\documentclass[aps,prb,twocolumn,showpacs,superscriptaddress,10pt]{revtex4-1}

\usepackage{graphicx}
\usepackage{amsmath,amssymb,amstext,amsthm}
\usepackage{bbold}
\usepackage{cancel}
\usepackage{color}
\usepackage{verbatim}
\usepackage{booktabs}
\usepackage[caption=false]{subfig}

\newcommand{\Z}{\mathbb{Z}}

\newcommand{\id}{\mathbb{1}}

\usepackage[utf8x]{inputenc}

\begin{document}

\title{Topological mirror insulators in one dimension}

\author{Alexander Lau}
\affiliation{Institute for Theoretical Solid State Physics, IFW Dresden, 
01171 Dresden, Germany}

\author{Jeroen van den Brink}
\affiliation{Institute for Theoretical Solid State Physics, IFW Dresden, 
01171 Dresden, Germany}
\affiliation{Institute for Theoretical Physics, TU Dresden, 01069 Dresden, Germany}
\affiliation{Department of Physics, Harvard University, Cambridge, Massachusetts 02138, USA}

\author{Carmine Ortix}
\affiliation{Institute for Theoretical Solid State Physics, IFW Dresden, 
01171 Dresden, Germany}
\affiliation{Institute for Theoretical Physics, Center for Extreme Matter and Emergent Phenomena, Utrecht University, Princetonplein 5, 3584 CC Utrecht, Netherlands}

\date{\today}

\begin{abstract}

We demonstrate the existence of topological insulators in one dimension
protected by mirror and time-reversal symmetries. 
They are characterized by a nontrivial $\Z_2$ topological invariant 
defined in terms of the ``partial"  polarizations,  which we show to 
be quantized in presence of a 1D mirror point. 
The topological invariant determines the generic presence or absence of 
integer boundary charges at the mirror-symmetric boundaries of the system. 
We check our findings against spin-orbit coupled Aubry-Andr\'e-Harper models
that can be realized, {\it e.g.} in cold-atomic Fermi gases loaded in
one-dimensional optical lattices or in density- and Rashba spin-orbit-modulated
semiconductor nanowires. 
In this setup, in-gap end-mode
Kramers doublets appearing in the topologically non-trivial state effectively
constitute
a double-quantum-dot with spin-orbit
coupling.   
 
\end{abstract}


\maketitle


\section{Introduction}

The study of topological phases of matter is one of the most active
research areas in physics. Triggered by the discovery of the quantum
Hall effect in 1980\cite{KDP80} and its theoretical explanation
in terms of the topological properties of the Landau levels,\cite{TKN82,Koh85}
a plethora of topologically nontrivial quantum
phases has been predicted theoretically and confirmed 
experimentally.\cite{BHZ06,KWB07,XQH09,RIR13,PRK15,KaM05,LFY11,LaT13,YZT12,
LZZ14,WTV11,HXB15}
In particular, topological insulators and superconductors
have been the subject of intensive research
efforts.\cite{Moo10,HaK10,QiZ11,Fu11} 
Unlike their
topologically trivial counterparts, they exhibit protected surface
or edge states inside the bulk excitation gap of the material.
These topological boundary states are the prime physical
consequence of the topology of the bulk band structure,
which is  encoded in  a quantized topological invariant.\cite{HaK10,RSF10}

Belonging to the symplectic class AII of the Altland-Zirnbauer (AZ)
classification, time-reversal 
invariant topological insulators in two (2D) and three dimensions (3D) are
characterized by a $\Z_2$ topological invariant. Primary examples of
topologically nontrivial systems include HgTe/CdTe quantum wells in
2D,\cite{KWB07} and  Bi$_2$Se$_3$ in 3D.\cite{XQH09} For one-dimensional (1D)
systems, instead, the AZ scheme does not predict a similar topological
invariant, and thus all insulating phases are expected to be topologically
equivalent. 

In the context of recent findings of novel topological states of matter
protected by additional continuous
symmetries,\cite{CYR13,ShS14,HPB11,LOB15_2,ZKM13,UYT13,SMJ12}
we show here that a crystalline symmetry, namely mirror
symmetry, leads to a class of 1D time-reversal invariant topological insulators 
beyond the standard AZ classification of topological insulators. 
Using the
concept of partial polarizations, we demonstrate that the concomitant presence
of mirror and time-reversal symmetry allows the definition of a $\Z_2$
topological invariant. 
A nonzero value of this invariant defines a one-dimensional
topological mirror insulator, which
hosts an odd number of integer electronic end charges.
We will present an explicit tight-binding model built upon the well-known 
Aubry-Andr\'e-Harper (AAH) model realizing this novel topological state of
matter. 
By calculating its topological invariants, we determine the
phase diagram of the system. Moreover, we consider energy spectra of
finite chains and calculate the corresponding values of the end charges to
explicitly confirm the
topological nature and the bulk-boundary correspondence.
We further study the effect of weak on-site disorder. We find that the
topological features are robust as long as mirror symmetry is preserved on
average. We finally show that a particular
realization of this model can be achieved in cold atomic Fermi gases loaded in
1D optical lattices and in semiconductor nanowires with gate-modulated Rashba
spin-orbit coupling. 


\section{$\Z_2$ Topological Invariant}

We start out by considering a generic system of spin one-half fermions in a 1D
crystalline potential with time-reversal symmetry $\Theta$. In addition, we will
consider the crystal to be mirror symmetric with respect to a 1D mirror point.
The Bloch Hamiltonian of such a system ${\cal H}(k)$, where $k\in(-\pi,\pi]$
with the lattice constant set to unity, then inherits the two following symmetry
constraints: 
\begin{eqnarray}
\Theta {\cal H}(k) \Theta^{-1} = {\cal H}(-k), \  \  {\cal M} {\cal H}(k){\cal
M}^{-1} = {\cal H}(-k),
\end{eqnarray}
where ${\Theta}=(\id \otimes i s^y)K$ is the antiunitary time-reversal operator,
while ${\cal M}={\cal I} \otimes i s^x$ is the unitary operator corresponding to
the operation of reflection with respect to the 1D mirror point. 
Without loss of generality, we assume the latter to be in the $\hat{x}$
direction. In addition, the 
$s^i$ are the usual Pauli matrices acting in spin space, while ${\cal I}$ acts
only on the spatial degrees of freedom and thus corresponds to spatial
inversion. 
For this representation of the symmetry operations, we have that $[\Theta,{\cal
M}]=0$, ${\Theta}^2=-1$ as required for spin one-half fermions, and ${\cal
M}^2=-1$, since spatial inversion must square to the identity, i.e., ${\cal
I}^2=1$. 

The Berry phase\cite{Zak89} $\gamma$ associated with the 1D Bloch Hamiltonian
${\cal H}(k)$ defines the charge polarization per unit cell\cite{FuK06} via 
$P_\rho = \gamma/2\pi$, with the electronic charge set to unity. Because of the
intrinsic $ 2\pi$ ambiguity of the Berry phase, $P_\rho$ is defined up to an
integer and can generally assume any value in between. However, this assertion
does not hold true for a 1D insulator with a 1D mirror point. To show this, we
first exploit the time-reversal symmetry of the system and consider the
\emph{partial polarizations}  introduced by Fu and Kane.\cite{FuK06} 
Due to Kramers' theorem every Bloch state at $k$ comes with a time-reversed
degenerate partner at $-k$. Hence, all occupied bands can be divided into two
time-reversed channels. The partial polarization $P^s$ is then just the charge
polarization of channel $s=\mathrm{I,II}$ 
such that $P_\rho=P^I + P^{II}$.
Using the symplectic time-reversal symmetry for spin one-half fermions, it is
possible to show that  $P^I=P^{II}$ modulo an integer [c.f.
Appendix~\ref{sec:quantization} and Ref.~\onlinecite{FuK06}].

The crux of the story is that in presence of mirror symmetry with the symmetry
operator ${\cal M}$ commuting with the time-reversal symmetry operator, the
partial polarizations 
satisfy $P^s=-P^s\: \mathrm{mod}\,1$ [c.f. Appendix~\ref{sec:quantization}].
Henceforth, $P^{s}$ can only assume the two distinct values $0$ and $1/2$
(modulo an integer).
The consequence of this result is twofold. First, 
it follows that the charge polarization $P_{\rho}$ of a 1D insulator with a 1D
mirror point is an integer quantity. 
Second, the quantized nature of the partial polarization allows to define a
$\Z_2$ topological invariant $\nu=2 P^{s} \mathrm{mod}\, 2 \equiv 0,1$. This
gives rise to two topologically distinct states that cannot be adiabatically
connected without closing the bulk energy gap or breaking the defining
symmetries. Using the $U(2N)$ invariant form of the partial
polarization,\cite{FuK06} where $2N$ is the number of occupied energy bands, the
topological invariant explicitly reads:
\begin{equation}
\nu := \frac{1}{\pi} \bigg[ \int_0^{\pi} dk\:\mathcal{A}(k) + i\,\log\Big(
\frac{\mathrm{Pf}[w(\pi)]}{\mathrm{Pf}[w(0)]} \Big)\bigg]
\,\mathrm{mod}\,2,
\label{eq:invariant}
\end{equation}
where $\mathcal{A}(k)=i\sum_{n\,\mathrm{occ.}} \langle
u_{k,n}|\partial_k|u_{k,n}\rangle$ is the 
Berry connection of all occupied bands and we introduced the $U(2N)$
matrix $w_{\mu\nu}(k)= \langle u_{-k,\mu}|\Theta|u_{k,\nu}\rangle$ which is
antisymmetric at $k=0,\pi$ and thus
characterized by its Pfaffian $\mathrm{Pf}(w)$.

The quantization of the partial polarization in one-dimensional systems with a mirror point has a direct physical consequence.
The charge polarization of a system is indeed directly connected to the accumulated
bound charge at its surfaces.\cite{KiV93} For a one-dimensional system the end
charge $Q$ is simply related to the polarization $P$ by $Q\,\mathrm{mod}\,1 =
P$. 
Since the partial polarization $P^\mathrm{I}$ is just the usual polarization
associated with one of the time-reversed channels, we can assign a
\emph{partial} bound charge to this channel, which is proportional to
$P^\mathrm{I}$. With two identical time-reversed
channels, the total bound charge per end is then
\begin{equation}
Q\,\mathrm{mod}\,2 = 2P_\mathrm{I} =
\nu.
\end{equation}

This establishes a direct connection between the
number of bound charges and the $\Z_2$ invariant $\nu$. In particular, 
systems for which $\nu=1$ are
topological mirror insulators: they are characterized by the presence 
of an \emph{odd} number of integer-valued electronic end charges at the mirror
symmetric boundaries of the system [c.f. Refs.~\onlinecite{KiV93,MOS16}]. 
This is the central result of this paper.

A few remarks are in order.
First, we emphasize that the
quantities in Eq.~\eqref{eq:invariant} require a continuous gauge. Such a gauge
can be straightforwardly constructed from numerically obtained eigenstates
[c.f. Appendix~\ref{sec:smooth_gauge}]. 
Second, we note that
the $\Z_2$ invariant of
Eq.~\ref{eq:invariant} cannot be determined from the knowledge of the electronic
wavefunctions only at the time-reversal invariant momenta, as it occurs for
crystalline topological insulators in 2D and 3D. Eq.~\ref{eq:invariant} requires
the knowledge of the wavefunctions in the \emph{entire} BZ. 
This also implies that for a topological phase transition, that occurs via a
closing and reopening of the 1D bulk gap at momentum $q$, $q$ is in general not
a high-symmetry point in the BZ.

We finally point out that the existence of our $\Z_2$ topological
invariant does not contradict the AZ classification of topological insulators
and superconductors,\cite{Zir96,AlZ97,HHZ05,SRF08,Kit09,RSF10}, which does not
take into account the point-group symmetries of the system. In the absence of
mirror symmetry, the partial polarizations of a 1D system with time-reversal
symmetry are indeed no longer quantized, and therefore $\nu$ does not represent
an invariant. The existence of 1D topological mirror insulators is instead in
agreement with the recent extensions of the original AZ classification taking
into account point-group symmetries,\cite{LuL14,CYR13,ShS14} in particular with
Refs.~\onlinecite{ShS14} and~\onlinecite{CYR13} which predict that
mirror-symmetric 1D systems in principle allow for a $\Z_2$
invariant. 
In this respect, it must be pointed out that certain types of
translational-symmetry breaking terms, such as charge-density
waves, can turn the system into a trivial insulator, similar to weak
topological insulators in 3D. From this point of view, a topological mirror
insulator can be viewed as a realization of a low-dimensional, \emph{weak}
topological insulator.

\begin{figure}[t]
\includegraphics[width=0.99\columnwidth]{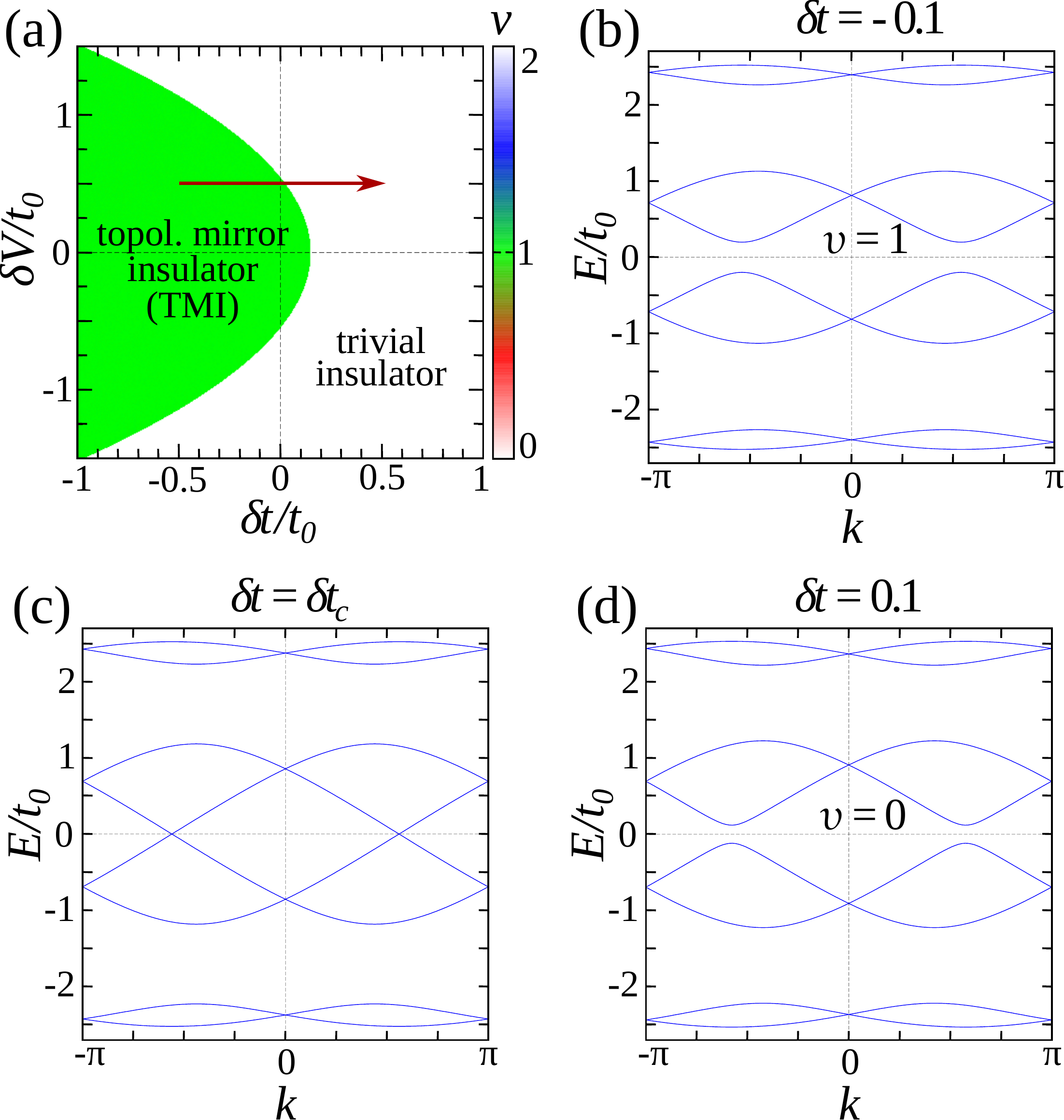}
\caption{(color online) Phase diagram and spectra of the spin-orbit coupled 
AAH model with $\alpha=\gamma=1/2$ (dimerization), $\beta=1/4$, $V_0=0$,
$\lambda_0=0.5t_0$, $\delta\lambda=-0.3t_0$, $\phi_t=\phi_\lambda=\pi$: (a)
half-filling phase diagram of the bulk for $\phi_V=-\pi/4$. The value of the
$\Z_2$ invariant $\nu$ is indicated by pixel color. Note that it only assumes
two distinct values, $0$ and $1$. (b-d) Bulk band structures for
$\delta V=0.5t_0$ and fixed $\delta t$ with (b) $\delta t=-0.1t_0$, (c)
$\delta t=\delta t_c\approx0.025$, 
(d) $\delta t=0.1t_0$. The band structures correspond to systems along the
arrow in (a). Note that the half-filling bulk energy gap closes away from the
time-reversal invariant points $k=0$ and $\pi$.}
\label{fig:AAH_phase_diagram_and_gap_closing}
\end{figure}


\section{Spin-orbit coupled Aubry-Andr\'e-Harper models}

We are now going to present an explicit model that features a 1D topological
mirror insulating phase. In particular, we will consider the
following tight-binding Hamiltonian for spin-$1/2$ electrons on a 1D lattice 
\begin{eqnarray}
\mathcal{H}&=&\sum_{j,\sigma} [t_0+\delta t\cos(2\pi\alpha j + \phi_t)]\, c_{j+1,\sigma}^\dagger c_{j\sigma}\nonumber\\
&&{}+ \sum_{j,\sigma} [V_0 + \delta V\cos(2\pi\beta j + \phi_V)]\, c_{j\sigma}^\dagger c_{j\sigma} \nonumber\\
&&{}+ i\sum_{j,\sigma,\sigma'} [\lambda_0 + \delta\lambda\cos(2\pi\gamma j + \phi_\lambda)]\, c_{j+1,\sigma}^\dagger s^y_{\sigma\sigma'}c_{j\sigma'} \nonumber\\
&&{}+ \mathrm{h.c.}
\label{eq:AAH_model_general}
\end{eqnarray}
This is a generalization of the famous Aubry-Andr\'e-Harper (AAH)
model.\cite{Har55,AuA80,GSS13,LOB15_2} It contains
harmonically modulated nearest-neighbor hopping, 
on-site potentials and spin-orbit coupling (SOC) with 
amplitudes $\delta t$, $\delta V$, $\delta\lambda$, phases $\phi_t$, $\phi_V$,
$\phi_\lambda$, and periodicities $1/\alpha$, $1/\beta$, $1/\gamma$. For
simplicity, we restrict the model to rational values of the periodicities.
Moreover, $t_0$, $V_0$ and $\lambda_0$ are the bond and site independent values
of hopping, potentials and SOC, the operators $c_{j\sigma}^\dagger$
($c_{j\sigma}$) create (annihilate) an electron with spin $\sigma$
($\sigma=\uparrow,\downarrow$)
at lattice site $j$, and the $s^i$ are Pauli matrices. 
The Hamiltonian of Eq.~\ref{eq:AAH_model_general} possesses time-reversal
symmetry whereas mirror symmetry is present only for specific values of the
phases  $\phi_t$, $\phi_V$ and $\phi_\lambda$. For the
computation of band structures and eigenstates we use exact numerical
diagonalization. 
In the case of periodic boundary conditions, we exploit the translational 
symmetry of the system and work with the corresponding Bloch Hamiltonian in
momentum space. For open boundary conditions, we take the real-space
Hamiltonian with a finite number of unit
cells.
The $\Z_2$ invariant of Eq.~\eqref{eq:invariant} is calculated numerically using 
the aforementioned procedure to construct a continuous $U(2N)$ gauge from
numerically obtained eigenstates [c.f. Appendix~\ref{sec:smooth_gauge}].

\begin{figure}[t]
\includegraphics[width=0.99\columnwidth]{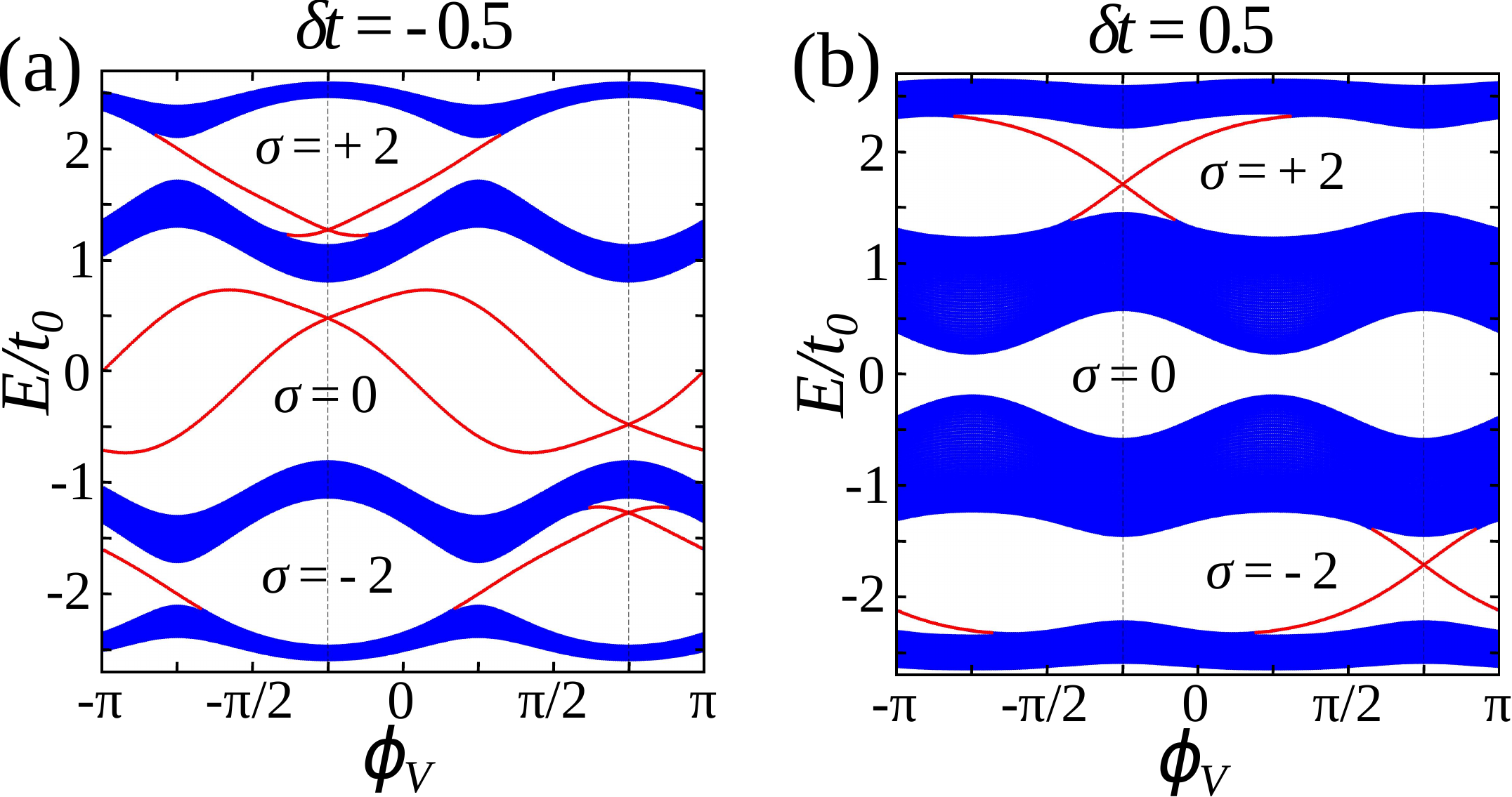}
\caption{(color online) Spectra of the spin-orbit coupled 
AAH model in a finite geometry for different $\phi_V$. Other parameters are
$\alpha=\gamma=1/2$ (dimerization), $\beta=1/4$, $V_0=0$, $\delta V=0.5t_0$
$\lambda_0=0.5t_0$, $\delta\lambda=-0.3t_0$, $\phi_t=\phi_\lambda=\pi$: 
(a) $\delta t=-0.5t_0$, (b) $\delta t=0.5t_0$. 
States localized at the ends of the chain are highlighted in red. In addition,
we display the Hall conductivities $\sigma$ in units of
$e^2/h$ associated with each gap when $\phi_V$ is interpreted as an additional
momentum variable.}
\label{fig:AAH_finite}
\end{figure}

We will consider the model in Eq.~\eqref{eq:AAH_model_general} with
$\alpha=\gamma=1/2$, $\phi_t=\phi_\lambda=\pi$, $V_0=0$, and $\beta=1/4$. 
With this choice of parameters, the unit cell of our model
Hamiltonian contains four lattice sites and the model preserves reflection
symmetry for $\phi_V=-\pi/4$ and $3\pi/4$.
This is a direct generalization of the model considered in
Ref.~\onlinecite{LOB15_2} to the spinful case. 
The model can be potentially realized using an ultracold Fermi gas loaded in 
a 1D optical lattice \cite{Bloch05,AAL13,CSH12,WYF12}, while the ensuing end
states
can be detected using time-of-flight measurements or optical
microscopy.\cite{MPC15,LGS16}
We emphasize that while state-of-the-art technologies allow to create SOC terms
explicitly breaking mirror symmetry,\cite{Zhai15,LLC13} the rapid progress in
the field can be expected to bring soon a much larger variety of SOC terms into
reach.\cite{Zhai15,CJS11,WKL15}

\subsection{Phase diagram and in-gap end states}

We first determine the phase diagram of our system with
respect to the $\Z_2$ invariant $\nu$ of
Eq.~\eqref{eq:invariant} for the half-filled system at $\phi_V=-\pi/4$, i.e.,
where the model has reflection symmetry. 
The phase diagram in the $\delta t$-$\delta V$ parameter space is shown in 
Fig.~\ref{fig:AAH_phase_diagram_and_gap_closing}(a).
We identify two phases which are
separated by a parabolic phase boundary: a trivial phase with $\nu=0$ on the
right side and a topological phase with $\nu=1$ on the left side. 
Similarly to the spinless version of the model,\cite{LOB15_2} the bulk 
energy gap at half-filling closes at the topological phase transition. 

Note that bulk band gap closing and reopening occurs
away from the time-reversal invariant momenta, as is shown in
Figs.~\ref{fig:AAH_phase_diagram_and_gap_closing}(b-d). This is in contrast to
other crystalline topological phases as pointed out in the
previous section. 

Moreover, the nontrivial topology of the model manifests itself through the
appearance of characteristic in-gap end states. This is checked and demonstrated
in Fig.~\ref{fig:AAH_finite} where we 
show the energy spectra of the AAH model under
consideration with a finite number of unit cells and open boundary conditions.
The phase shift
$\phi_V$ of the on-site modulation is varied smoothly from $-\pi$ to $\pi$
thereby passing through the mirror-symmetric cases at $-\pi/4$ and $3\pi/4$.
At these points, we find four degenerate in-gap end states at half filling provided we
are
in the nontrivial area of the phase diagram of Fig.~\ref{fig:AAH_finite}(a).
Away from the mirror-symmetric points the observed states are split into two
degenerate pairs.
The two-fold degeneracy remains
since the model in Eq.~\ref{eq:AAH_model_general} preserves time-reversal
symmetry at all values of $\phi_V$. 
The half-filling end states are not encountered for parameters of the model for 
which we are in the trivial region of the phase diagram, as can be seen in
Fig.~\ref{fig:AAH_finite}(b).

Taking a different perspective, the appearance of end states at the $1/4$ and
$3/4$ filling gaps can also be understood by interpreting the phase $\phi_V$ as
the momentum of an additional artificial dimension. In this case, our model of
Eq.~\ref{eq:AAH_model_general} can be mapped to a dimerized Hofstadter
model\cite{Hof76,GSS13,LOB15_2,MCO15} for spinfull fermions with SOC in one
direction only. Contrary to models investigated before,\cite{COR12,OCR13} the
resulting model explicitly breaks the 2D time-reversal symmetry constraint
$\Theta^{-1} {\cal H}(k ,\phi_V) \Theta \neq {\cal H}(-k, -\phi_V)$, thereby
allowing for insulating states with nonzero Chern
numbers.\cite{LOB15_2,TKN82,Koh85} By calculating the Hall conductivities, we
find that they are doubled with respect to the conventional spinless Hofstadter
model\cite{LOB15_2} indicating that the two spin channels of our model carry
the \emph{same} topological content. This, in turn, implies that the in-gap
states
at $1/4$ and $3/4$ filling, appearing in Fig.~\ref{fig:AAH_finite},
correspond to the chiral edge states of a
generalized Hofstadter model in a ribbon geometry.
Furthermore, they correspond to insulating states with Hall conductivities $\pm
2$. On the contrary, the insulating phase at half-filling has zero Hall
conductivity but displays two quartets of
edge states in the topological phase originating from the 1D
mirror-symmetric cuts [c.f. Fig.~\ref{fig:AAH_finite}(a)].
We point out that
these results, as a generalization of                      
Ref.~\onlinecite{LOB15_2}, extend to arbitrary rational $\beta=p/q$ with $p$
and $q$ coprime.

\subsection{Bulk-edge correspondence}

We have shown that the appearance of electronic in-gap end states is
characteristic of a topological mirror insulator. However, due to the absence of
chiral symmetry, which would pin the end modes at the center of the gap, 
symmetry-allowed perturbations can push the end
modes into the continuum of bulk states. Such perturbations could, for
instance, be introduced by surface potentials.

\begin{figure}[t]
\centering
\includegraphics[width=\columnwidth]{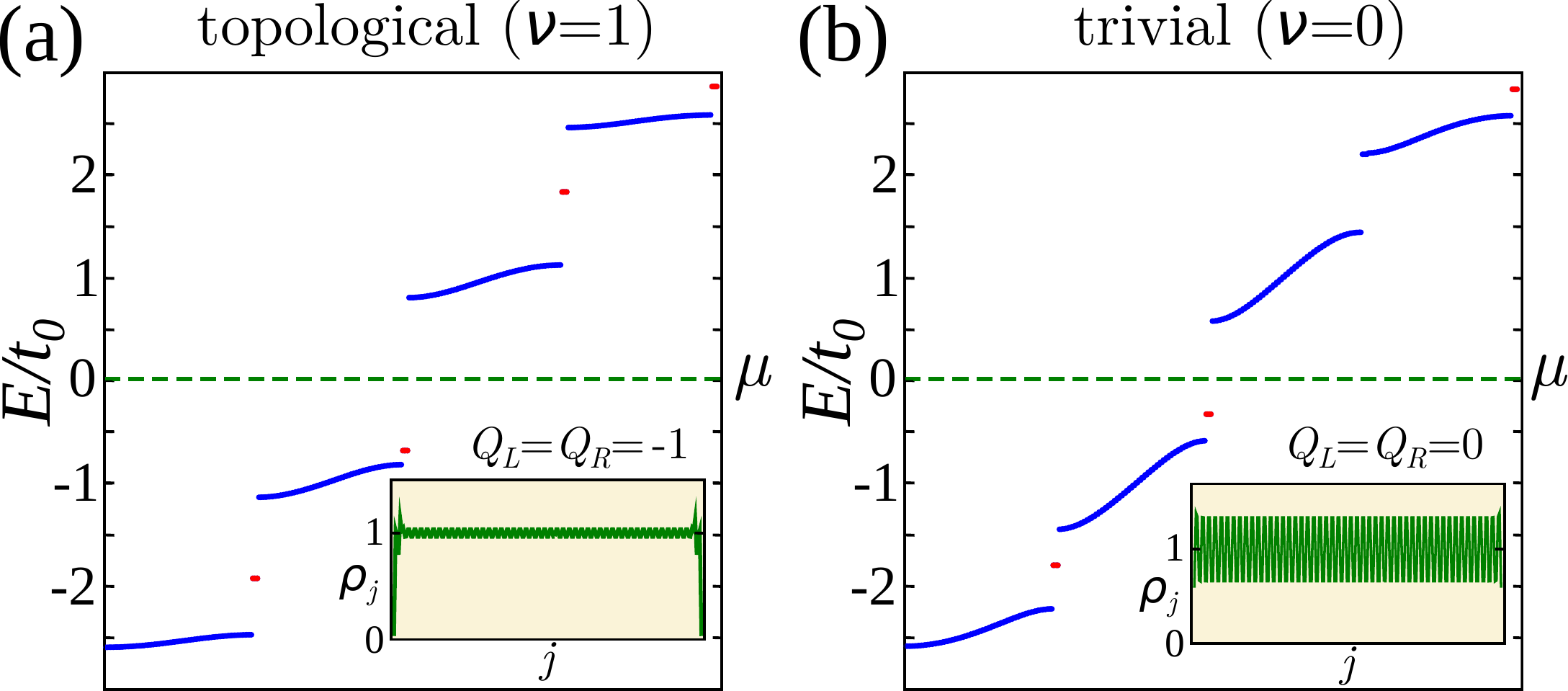}
\caption{(color online) Spectra and local charge densities of the
mirror-symmetric, spin-orbit coupled AAH model with open
boundary conditions and $\alpha=\gamma=1/2$, $\beta=1/4$,
$V_0=0$, $\delta V=0.5t_0$, $\lambda_0=0.5t_0$, $\delta\lambda=-0.3t_0$,
$\phi_t=\phi_\lambda=\pi$, $\phi_V=-\pi/4$, and a surface potential
$V_\mathrm{LR}=0.6t_0$:
(a) Topological phase with $\delta
t=-0.5t_0$,
(b) trivial phase with $\delta
t=0.4t_0$.
The main panels show the energy spectra with end states highlighted in red. The
dashed line signifies the chemical potential $\mu$ used for the
calculation of the local charge densities $\rho_j$. The latter are presented in
the insets. In addition, we display the corresponding values of the electronic
end charges $Q_\mathrm{L}$ and $Q_\mathrm{R}$.}
\label{fig:bands_with_surface_potential}
\end{figure}

We explicitly demonstrate the point above by adding a generic surface
potential $\sum_\sigma V_\mathrm{LR}(c_{1\sigma}^\dagger c_{1\sigma} + c_{L\sigma}^\dagger
c_{L\sigma})$ to our model with open boundary conditions, and
analyzing the ensuing energy spectra. Again, we fix the on-site potential
phase $\phi_V$ to $-\pi/4$ such that our model preserves mirror symmetry.

Without the surface potential, we observed 
two degenerate in-gap Kramers pairs at half filling in the topological phase
[see Fig.~\ref{fig:AAH_finite}(a)]. This picture changes if we switch on the
surface potential $V_\mathrm{LR}$ [see
Fig.~\ref{fig:bands_with_surface_potential}(a)]. We observe that the end modes
in the half-filling gap are pushed up into the
conduction band, while another degenerate pair of Kramers
doublets emerges from the valence band. However, the
appearance of these states cannot be linked to the topological invariant of the
system. In fact, these additional states occur both in the
topological and in the trivial phases, as can be seen by 
comparing Figs.~\ref{fig:bands_with_surface_potential}(a) and~(b) at
half-filling.

Having established that sufficiently strong 
symmetry-allowed edge potentials are detrimental for the occurrence of in-gap
end modes, we now proceed to uncover the bulk-edge correspondence for 1D
topological mirror insulators, {\it i.e.}, the
existence of an odd number of integer-valued electronic end charges.

We define the \emph{end charge} of a system to be 
the net deviation of the local charge density close to the end from the average
charge density in the bulk. Adopting the definition of Ref.~\onlinecite{PYK16},
we calculate the \emph{left} end charge $Q_\mathrm{L}$ as the limit of
\begin{equation}
\sum_{j}^{L} \Theta(l_0-j)(\rho_j-\bar{\rho})
\end{equation}
for sufficiently large $l_0$. A similar definition is used for
the \emph{right} end charge $Q_\mathrm{R}$ with $\Theta(l_0-L+j)$. Here, $L$ is
the length of the chain, $\Theta(x)$ is the Heaviside function and $l_0$ is a
cut-off. $\rho_j=\sum_\nu^N|\psi_\nu(j)|^2$ is the local charge density of the
ground state in units of $-e$ with the sum running over all $N$ occupied states
$\psi_\nu$ up to the chemical potential $\mu$. The bulk charge density
$\bar{\rho}$ is treated as a constant background that is fixed by the chemical
potential. In particular, at half filling we have $\bar{\rho}=1$ corresponding
to one electron per site. 

The insets in Fig.~\ref{fig:bands_with_surface_potential} show the local
charge densities of the finite AAH chains when the chemical potential $\mu$ is
in the half-filling bulk energy gap above or below the end states. 
As expected, the local charge density
oscillates around $1$ in the bulk. Moreover, in the topological phase there are
large deviations from this value near the ends of the system indicating the
presence of electronic end charges. On the contrary, there are no pronounced
features in the trivial phase.

This is confirmed by explicit calculation of the end charges. In the
topological phase [see Fig.~\ref{fig:bands_with_surface_potential}(a)] we find
$Q_\mathrm{L}=Q_\mathrm{R}=+1$ when $\mu$ is placed right above the topological
end states. Leaving the \emph{two} boundary states at each end unoccupied
leads to an end charge of $-1$. Hence, there is a direct correspondence between
end states and end charges. End states always come in degenerate pairs due
to time-reversal symmetry. From this we see that, regardless of where we put the
chemical potential in the bulk energy gap and regardless of how many pairs of
degenerate Kramers pairs are present, the end charge will always assume an
\emph{odd} value in agreement with the general analysis of the previous
section. More importantly, this is a robust feature that is not
affected by the presence of surface potentials.

In contrast to that, in the trivial
phase [see Fig.~\ref{fig:bands_with_surface_potential}(b)] we find
$Q_\mathrm{L}=Q_\mathrm{R}=0$ for a chemical potential above the trivial end
states. Hence, only \emph{even} values of end charges are possible. For
instance, with the trivial states unoccupied we calculate end charge values of
$-2$. 
In addition,
the end charges no longer assume integer values if
mirror symmetry is broken.

\subsection{Effect of disorder}

\begin{figure}[t]
\centering
\includegraphics[width=\columnwidth]{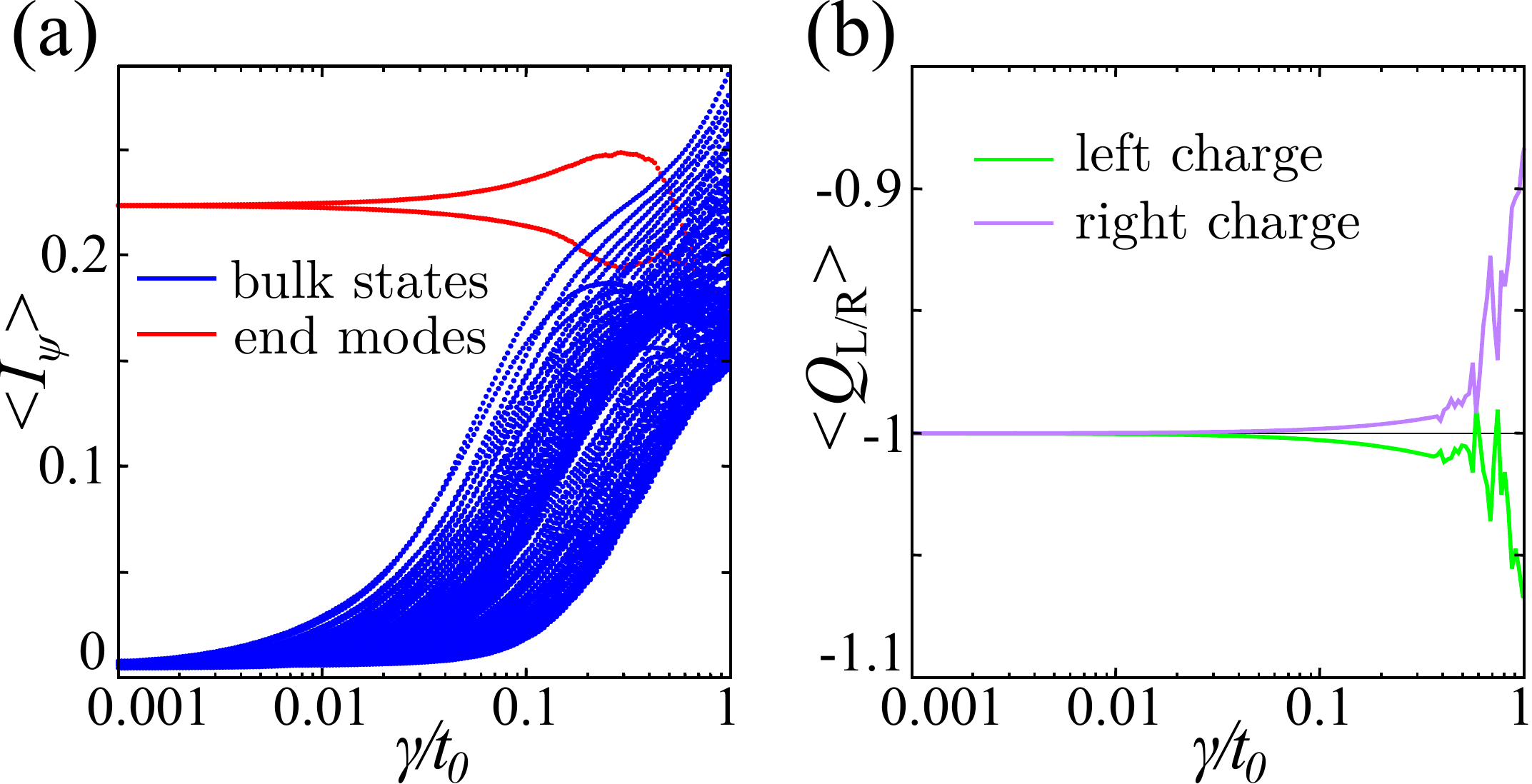}
\caption{(color online) Effect of disorder in the half-filled, spinful AAH model
with open boundary conditions and $\alpha=\gamma=1/2$,
$\beta=1/4$, $t=-0.5t_0$ $V_0=0$, $\delta V=0.5t_0$, $\lambda_0=0.5t_0$,
$\delta\lambda=-0.3t_0$, $\phi_t=\phi_\lambda=\pi$, and $\phi_V=-\pi/4$
(topological phase): (a) Disorder-averaged IPR $\langle I_\psi\rangle$ for bulk
states and end states. (b) Disorder-averaged end charges
$\langle Q_\mathrm{L}\rangle$ and $\langle Q_\mathrm{R}\rangle$
with the
chemical potential right below the in-gap end states. Shown is the
dependence on the standard deviation $\gamma$ of the random disorder potential.}
\label{fig:disorder}
\end{figure}

We have considered the fate of the characteristic in-gap end states and of the
topological end charges of a 1D
mirror insulator under the influence of disorder. 
The topological nature of a 1D topological mirror insulator is
protected by time-reversal and mirror symmetry. The latter is a spatial
symmetry and is, in general, broken by disorder. However, recent studies on
topological crystalline insulators have shown that their topological features
are still present if the protecting symmetry is preserved on
average.\cite{ShS14} We will now demonstrate that this holds also for our class
of systems. To show this, we model the the effect of disorder
by adding a random nonmagnetic on-site potential to our model of the
form $\sum_j W_j c_{j\sigma}^\dagger c_{j\sigma}$, where the $W_j$ are
independent random variables subject to a Gaussian distribution with zero mean
and standard deviation $\gamma$. The latter is a measure for the disorder
strength.

To analyze the effect of disorder on the end states we consider
the expectation value of the inverse participation ratio (IPR), which is a
quantitative measure of localization.\cite{PYK16,PaO16} The IPR of a given
state $\psi$ is defined
as $I_\psi = \sum_j^L |\psi(j)|^4$ with $|\psi(j)|^2$ being the weight of the
state at site $j$. The IPR assumes values in the interval $(0,1]$. An IPR
of $1$ corresponds to a perfectly localized state, whereas small values indicate
a state equally distributed over the whole length of the system.

In Fig.~\ref{fig:disorder}(a) we present the IPR of
occupied states for the finite, half-filled AAH chain of length $L=200$ averaged
over $10^3$ random disorder configurations. The model
parameters are chosen such that the disorder-free chain preserves mirror
symmetry and is in the topological phase. We
observe 
that the IPR of the topological end states stays nearly unaffected at a large
value as long as the disorder is weak ($\gamma\lesssim 0.01t_0$). In contrast to
that, the IPR of the bulk states is more than one order of magnitude lower. For
stronger disorder the topological end modes are still well-localized, but their
IPR starts to deviate from their previously constant value due to mixing with
bulk states. However, this does not lead to a sizable decrease of the IPR due to
the onset of Anderson localization. The latter also accounts for the substantial
increase in the IPR of the bulk states.

In addition, in Fig.~\ref{fig:disorder}(b) we show the disorder-averaged
values of the boundary charges $Q_\mathrm{L}$ and $Q_\mathrm{R}$ in the same
setting. The chemical
potential, which determines the number of occupied states, is set to be in the
half-filling bulk energy gap such that the topological end states are
unoccupied. In the previous section, we saw that the disorder-free values of
the end charges are exactly $-1$. In the presence of disorder this value
is barely affected up to intermediate disorder strength
($\gamma\lesssim
0.1t_0$). Only for strong disorder we see considerable deviations. 

In the presence of nonmagnetic
disorder with zero mean the
characteristic end states of a topological mirror insulator remain, in conclusion,
well-localized and its topological end charges remain sharply quantized.

\begin{figure}[t]\centering
\includegraphics[width=.8\columnwidth]{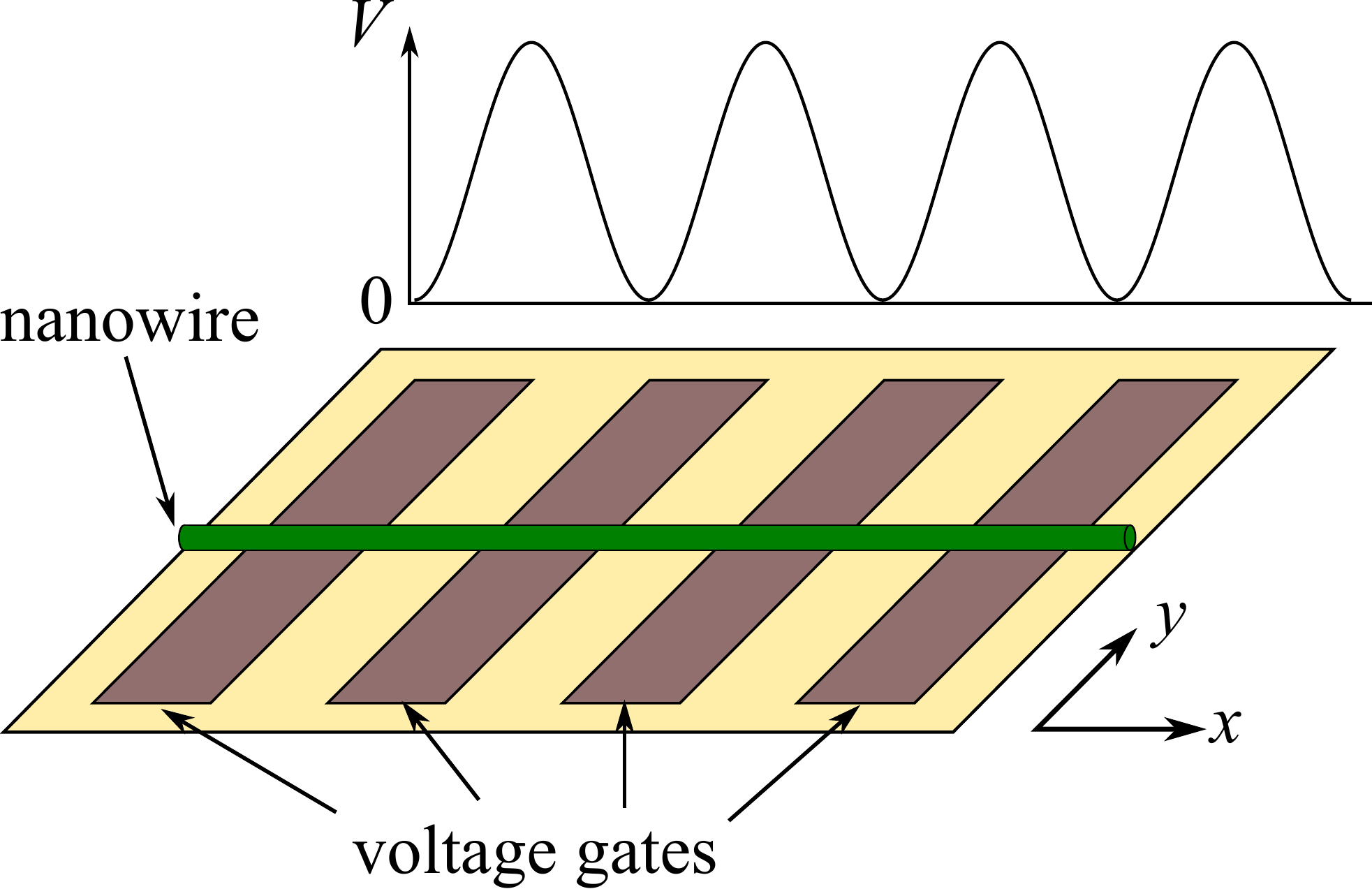}
\caption{(color online) A semiconductor nanowire setup with perpendicular 
modulated voltage gates. The gates induce a modulation of both on-site
potentials
and Rashba SOC. Subject to mirror symmetry, such a nanowire hosts
localized in-gap Kramers pairs at its ends.}
\label{fig:nanowire}
\end{figure}


\section{Density- and Rashba spin-orbit-modulated semiconductor nanowires}

We finally show that the general model of Eq.~\ref{eq:AAH_model_general} allows 
for topological mirror insulating phases in a large portion of its parameter
space. To demonstrate this, we consider our model with constant hopping
parameters  ($\delta t=0$), larger but equal periods of on-site potentials and
SOC ($\beta=\gamma$), and nonzero average values $t_0$, $V_0$, and $\lambda_0$.
In this parameter regime the model corresponds to the tight-binding Hamiltonian
for a semiconductor nanowire with Rashba SOC where opportunely designed finger
gates can cause a periodic density modulation as well as a gate-tuned modulation
of the Rashba SOC strength. This is illustrated in Fig.~\ref{fig:nanowire}. 
In this setup, end states can be detected using tunneling density of states.

\begin{figure}[t]\centering
\includegraphics[width=\columnwidth]{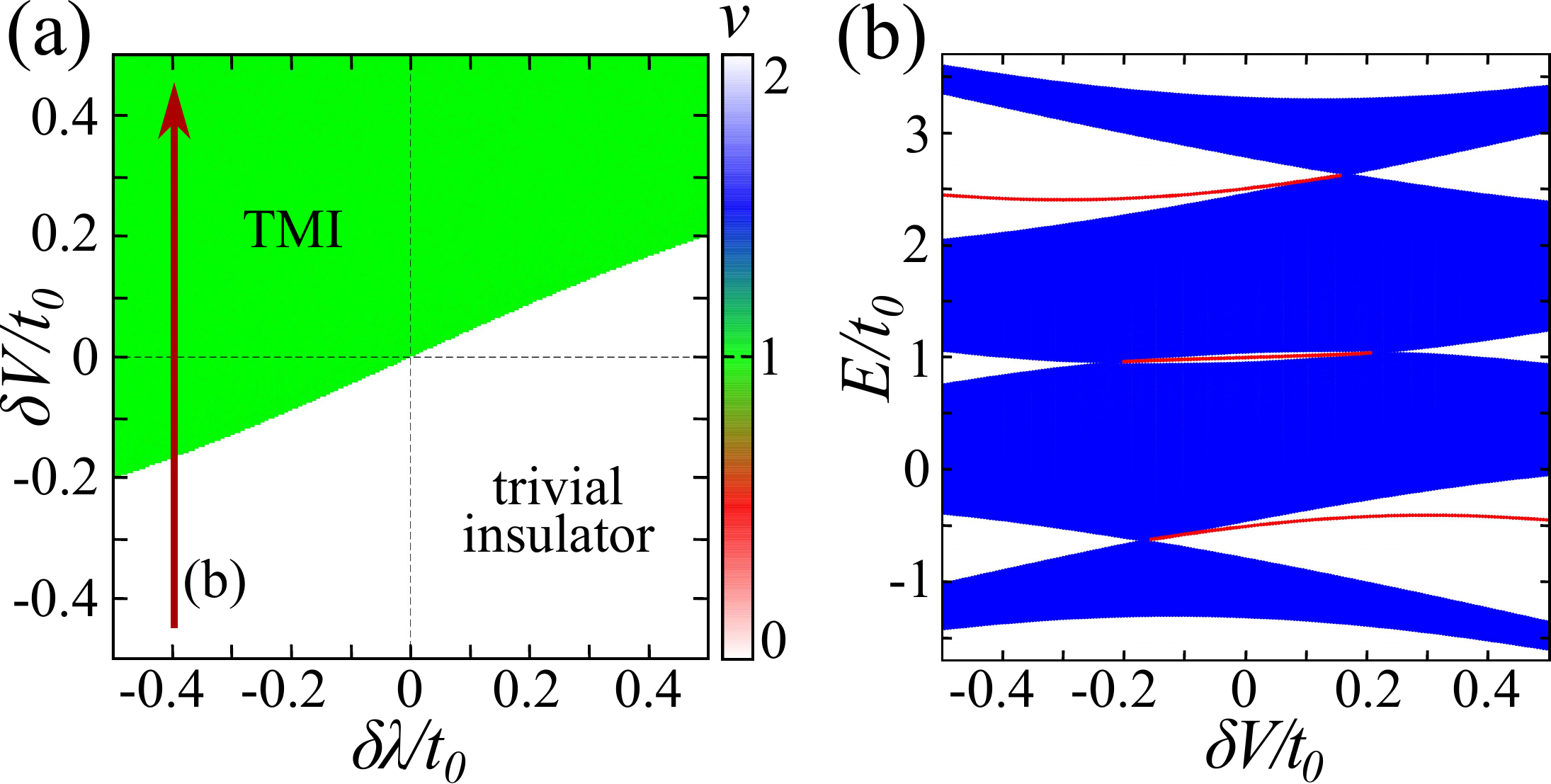}
\caption{(color online) Phase diagram and spectra of a density- and Rashba 
spin-orbit-modulated semiconductor nanowire with $\beta=\gamma=1/4$,
$V_0=0.5t_0$, $\lambda_0=0.5t_0$, $\delta t=0$, $\phi_V=3\pi/4$, and
$\phi_\lambda=\pi$: (a) Quarter-filling phase diagram. The value of the $\Z_2$
invariant $\nu$ is indicated by pixel color. The arrow shows a path through the
phase diagram. (b) Energy spectra of nanowires with open boundary conditions at
fixed $\delta\lambda=-0.4t_0$ corresponding to $\delta V$ values along the path
shown in (a). States localized at the end points of the wires are highlighted in
red. Note that there are end states at filling fractions $1/4$, $1/2$, and $3/4$
in certain ranges of $\delta V$.}
\label{fig:Rashba4_phase_diagram_and_finite}
\end{figure}

In Fig.~\ref{fig:Rashba4_phase_diagram_and_finite}(a), we show a
quarter-filling 
$\delta V$-$\delta\lambda$ phase diagram with respect to the $\Z_2$ invariant of
Eq.~\eqref{eq:invariant} for $\beta=\gamma=1/4$, where the corresponding unit
cell comprises four lattice sites. The modulation phases are chosen such that
the model respects mirror symmetry. The Rashba term preserves this symmetry if
$\phi_{\lambda,m}=0$ or $\pi$, whereas the on-site term is reflection symmetric
for $\phi_{V,m}=-\beta\pi$ or $(1-\beta)\pi$.\cite{LOB15_2} Again, we identify
two distinct topological phases. The shape of the phase boundary is mainly
influenced by the relative magnitude of $\lambda_0$ and $V_0$. 
Moreover, the phase with nontrivial $\Z_2$ invariant features
characteristic 
Kramers pairs at the end points of the corresponding wire.
This is demonstrated
in Fig.~\ref{fig:Rashba4_phase_diagram_and_finite}(b) for a path through the
phase diagram at constant $\delta\lambda$ and variable $\delta V$. At quarter
filling, end states are absent for small values of $\delta V$. However, if we
increase $\delta V$ the bulk energy gap closes and reopens again, revealing two
degenerate Kramers pairs localized at the end points of the wire. We also
observe localized Kramers pairs at filling fractions $1/2$ and $3/4$ that can be
attributed to a nontrivial $\Z_2$ invariant. 

In addition, we calculate end charge values of $\pm 1$ for the topological
phases at $1/4$, $1/2$, and $3/4$ filling depending on whether the chemical
potential is tuned above or below the degenerate in-gap Kramers pairs. On the
contrary, the trivial phases exhibit no end charges.
This once again verifies the
bulk-boundary correspondence. 

In general, the observed end states render an effective spin-orbit coupled
quantum-dot system 
which can potentially be implemented to realize spin-orbit
qubits.\cite{NFB10,LiY14} We point out that the use of finger gates in a
realistic system is expected to produce an equal phase modulation of the Rashba
SOC and onsite potential  ($\phi_V=\phi_\lambda\equiv\phi$) which breaks the
mirror symmetry in our tight-binding description. However, in the $\beta
\rightarrow 0$ continuum limit one has $\phi_{V,m}\rightarrow \phi_{\lambda,m}$,
which shows that a density- and Rashba SOC-modulated mirror-symmetric
semiconductor nanowire can be realized in practice. 


\section{Conclusions}

To conclude, we have shown that one-dimensional spin-$1/2$ fermionic 
systems with both time-reversal and mirror symmetry can have nontrivial
topology. In particular, we have found that the partial polarization in these
systems can only assume the values $0$ or $1/2$. The partial polarization can
therefore be used as a topological $\Z_2$ invariant. If this number is nonzero,
the system is a topological mirror insulator whose hallmark is an odd number
of electronic end charges. Furthermore, these end charges are robust against
weak disorder as long as the protecting mirror symmetry is preserved on
average.
We have checked these findings against a class of models,
corresponding to generalized AAH models with SOC that realize
topologically nontrivial $\Z_2$ phases.
These models can be realized in Fermi gases loaded in periodical optical
lattices, 
as well as in semiconductor nanowires with perpendicular modulated voltage
gates. In the latter setup the characteristic pairs of in-gap end states realize an
effective spin-orbit coupled double-quantum-dot system potentially relevant for
spin-orbit qubits. 

We thank Guido van Miert for stimulating discussions, and acknowledge the
financial support of the Future and Emerging Technologies (FET) programme within
the Seventh Framework Programme for Research of the European Commission 
under FET-Open grant number: 618083 (CNTQC).  This work has been 
supported by the Deutsche Forschungsgemeinschaft under Grant No. OR 404/1-1 and
SFB 1143.  JvdB acknowledges support from the Harvard-MIT Center for Ultracold
Atoms.

\newpage
\appendix

\section{Quantized partial polarization}
\label{sec:quantization}

In the following, we briefly review the concept of partial polarization and show
that its value is quantized in the presence of reflection
symmetry. 

Let us start with the general charge polarization $P_\rho$. It is a measure for
the electric dipole moment per unit cell and can be elegantly written in terms
of the Berry connection\cite{Zak89,KiV93,Res94,Res00}:
\begin{equation}
P_\rho = \frac{1}{2\pi}\int_{-\pi/a}^{\pi/a} dk\: \mathcal{A}(k),
\label{eq:polarization}
\end{equation}
with the $U(1)$ Berry's connection
\begin{equation}
\mathcal{A}(k)=i\sum_n \langle u_{k,n}|\partial_k|u_{k,n}\rangle.
\label{eq:full_berry_connection}
\end{equation}
Here, $|u_{k,n}\rangle$ is the lattice-periodic part of a Bloch state at
momentum $k$ and band index $n$, $a$ is the lattice constant, $q=-e$ is the
electron charge, and the sum is over all occupied bands. Since the right-hand
side of Eq.~\eqref{eq:polarization} is equivalent to the famous Berry
phase~\cite{Zak89} times a factor, $P_\rho$ is in general only defined up to an
integer and can assume any value in between. 

In the context of time-reversal invariant topological insulators, Fu and Kane
introduced the notion of \emph{partial polarization}~\cite{FuK06}. For this,
they made use of the well-known Kramers' theorem:  for a time-reversal symmetric
system with half-integer total spin, every energy level is evenly degenerate.
For a translationally invariant system this is equivalent to saying that every
Bloch state at $k$ comes with a time-reversed degenerate partner at $-k$. In
particular, states at the time-reversal invariant momenta $k=0$ and $\pi/a$ must
be evenly degenerate. Hence, a fully gapped system must have an even number of
occupied energy bands. Assuming, for simplicity, there are no other degeneracies
than those required by time-reversal symmetry, we can divide the $2N$ occupied
bands into $N$ pairs subject to~\cite{FuK06}
\begin{equation}
|u_{-k,\alpha}^\mathrm{I}\rangle = -e^{i\chi_\alpha(k)}\, \Theta
|u_{k,\alpha}^{\mathrm{II}}\rangle,
\label{eq:I_II_partition}
\end{equation}
where $\Theta$ is the antiunitary time-reversal operator with $\Theta^2=-1$,
$\alpha=1,\ldots,N$, and I,~II are the two time-reversed channels. Then, the
partial polarizations are simply the polarizations associated with the two
channels, i.e.,
\begin{equation}
P^s = \frac{1}{2\pi}\int_{-\pi/a}^{\pi/a} dk\: \mathcal{A}^{s}(k),\:\:
s=\mathrm{I,II},
\label{eq:def_partial_pol}
\end{equation}
where $\mathcal{A}^{s}(k)=i\sum_\alpha \langle
u_{k,\alpha}^s|\partial_k|u_{k,\alpha}^s\rangle$. 
It is sufficient to consider $P^\mathrm{I}$ only, because
\begin{eqnarray}
P^\mathrm{I} \!-\! P^\mathrm{II} &=& \frac{1}{2\pi} \int_{-\pi/a}^{\pi/a}
dk\:[\mathcal{A}^\mathrm{I}(k)-\mathcal{A}^\mathrm{II}(k)] \nonumber\\
&=& \frac{1}{2\pi} \int_0^{\pi/a}\!\! dk\:[\mathcal{A}^\mathrm{I}(k) \!+\!
\mathcal{A}^\mathrm{I}(-k) \!-\! \mathcal{A}^\mathrm{II}(k) \!-\!
\mathcal{A}^\mathrm{II}(-k)] \nonumber\\
&=& \frac{1}{2\pi}\sum_\alpha \int_0^{\pi/a} dk\:[\partial_k \chi_\alpha(k) -
\partial_k \chi_\alpha(-k)] \nonumber\\
&=& \frac{1}{2\pi} \underbrace{\sum_\alpha [\chi_\alpha(\pi/a) -
\chi_\alpha(-\pi/a)]}_{=2\pi m,\: m\in\Z} = m\in\Z,
\end{eqnarray}
where we have used that
\begin{eqnarray}
\mathcal{A}^\mathrm{I}(-k) &=&  i\sum_\alpha \langle
u_{-k,\alpha}^\mathrm{I}|\partial_{-k}|u_{-k,\alpha}^\mathrm{I}\rangle
\nonumber\\
&=& -i\sum_\alpha e^{-i\chi_\alpha(k)}\langle \Theta
u_{k,\alpha}^\mathrm{II}|\partial_{k}|e^{i\chi_\alpha(k)}\Theta
u_{k,\alpha}^\mathrm{II}\rangle\nonumber\\
&=& -i\sum_\alpha \langle \Theta u_{k,\alpha}^\mathrm{II}|\partial_{k}\Theta
u_{k,\alpha}^\mathrm{II}\rangle + \sum_\alpha
\partial_k\chi_\alpha(k)\nonumber\\
&=& -i\sum_\alpha \langle \Theta^2\partial_k u_{k,\alpha}^\mathrm{II}|\Theta^2
u_{k,\alpha}^\mathrm{II}\rangle + \sum_\alpha
\partial_k\chi_\alpha(k)\nonumber\\
&=& -i\sum_\alpha \langle \partial_k
u_{k,\alpha}^\mathrm{II}|u_{k,\alpha}^\mathrm{II}\rangle + \sum_\alpha
\partial_k\chi_\alpha(k)\nonumber\\
&=& \mathcal{A}^\mathrm{II}(k) + \sum_\alpha \partial_k\chi_\alpha(k),
\end{eqnarray}
with the properties $\langle \Theta v|\Theta w\rangle = \langle w|v\rangle$ and
$\Theta ^2=-1$ of the antiunitary time-reversal operator.
Hence, we have $P^\mathrm{II} = P^\mathrm{I}\,\mathrm{mod}\,1$.

On the other side, if the system preserves mirror symmetry with
\begin{equation}
\mathcal{M}\mathcal{H}(k)\mathcal{M}^{-1} = \mathcal{H}(-k)\: \textrm{ and }\:
[\mathcal{M},\Theta]=0,  
\end{equation}
where $\mathcal{H}(k)$ is the Bloch Hamiltonian of the system and $\mathcal{M}$
is the reflection operator, we further have $P_\mathrm{I} = -P_\mathrm{I}\,
\mathrm{mod}\,1$. This can be seen as follows:
assume again for simplicity that we have no other degeneracies than those
required by time-reversal symmetry. Then, we can write, similar to
Eq.~\eqref{eq:I_II_partition},
\begin{equation}
|\tilde{u}^\mathrm{I}_\alpha(-k)\rangle :=
-e^{i\beta(k)}\mathcal{M}|u^\mathrm{II}_\alpha(k)\rangle,
\label{eq:M_uII_relation}
\end{equation}
where $|\tilde{u}^\mathrm{I}_\alpha(k)\rangle$ is an eigenstate of the
Hamiltonian. Moreover, we can always choose $|u^\mathrm{II}_\alpha(k)\rangle$
such that $|\tilde{u}^\mathrm{I}_\alpha(k)\rangle$ is equal to 
$|u^\mathrm{I}_\alpha(k)\rangle$ up to a phase,
\begin{equation}
|\tilde{u}^\mathrm{I}_\alpha(k)\rangle =
e^{i\lambda(k)}|u^\mathrm{I}_\alpha(k)\rangle.
\label{eq:utilde_u_relation_1}
\end{equation}

Then, using Eqs.~\eqref{eq:I_II_partition},~\eqref{eq:M_uII_relation} and
$[\mathcal{M},\Theta]=0$, we easily see that
\begin{equation}
|u^\mathrm{I}_\alpha(k)\rangle =
-e^{i[\chi(-k)+\beta(-k)]}\mathcal{M}\Theta|\tilde{u}^\mathrm{I}
_\alpha(k)\rangle
\label{eq:utilde_u_relation_2}
\end{equation}
With this, we get
\begin{eqnarray}
P^\mathrm{I} &=& \frac{i}{2\pi}\sum_\alpha \int_{-\pi/a}^{\pi/a} dk\: \langle
u_{k,\alpha}^\mathrm{I}|\partial_k|u_{k,\alpha}^\mathrm{I}\rangle\nonumber\\
&\overset{\eqref{eq:utilde_u_relation_2}}{=}& \frac{i}{2\pi} \sum_\alpha
\int_{-\pi/a}^{\pi/a} dk\: \langle
\mathcal{M}\Theta\tilde{u}_{k,\alpha}^\mathrm{I}|\partial_k
\mathcal{M}\Theta|\tilde{u}_{k,\alpha}^\mathrm{I}\rangle\nonumber\\
&&- \frac{1}{2\pi} \underbrace{\sum_\alpha \int_{-\pi/a}^{\pi/a} \!\!\!dk\:
\partial_k [\chi(-k)\!+\!\beta(-k)]}_{= 2\pi n,\: n\in\Z}.
\end{eqnarray}
Since $P^\mathrm{I}$ is defined only up to an integer, we drop the
second term and continue as follows
\begin{eqnarray}
P^\mathrm{I} &=& \frac{i}{2\pi} \sum_\alpha \int_{-\pi/a}^{\pi/a} dk\: \langle
(\mathcal{M}\Theta)^\dagger\partial_k \mathcal{M}\Theta
\tilde{u}_{k,\alpha}^\mathrm{I}|\tilde{u}_{k,\alpha}^\mathrm{I}
\rangle\nonumber\\
&=& -\frac{i}{2\pi} \sum_\alpha \int_{-\pi/a}^{\pi/a} dk\: \langle
\tilde{u}_{k,\alpha}^\mathrm{I}|\partial_k|\tilde{u}_{k,\alpha}^\mathrm{I}
\rangle
\nonumber\\
&\overset{\eqref{eq:utilde_u_relation_1}}{=}& -\frac{i}{2\pi} \sum_\alpha
\int_{-\pi/a}^{\pi/a} dk\: \langle
u_{k,\alpha}^\mathrm{I}|e^{-i\lambda(k)}\partial_k
e^{i\lambda(k)}|u_{k,\alpha}^\mathrm{I}\rangle\nonumber\\
&=& -P^\mathrm{I} + m,\: m\in\Z,
\end{eqnarray}
where we have used that $\mathcal{M}\Theta$ is antiunitary with properties
$\langle(\mathcal{M}\Theta)^\dagger v|w\rangle = \langle \mathcal{M}\Theta
w|v\rangle$ and $(\mathcal{M} \Theta)^\dagger \mathcal{M}\Theta=1$.

In conclusion, 
we have shown that $P^\mathrm{I} = -P^\mathrm{I}\, \mathrm{mod}\,1$. This means
that $P^I$ can only
assume the values $0$ or $1/2$ modulo $1$ and is therefore quantized.

\section{Smooth $U(N)$ gauge for 1D systems}
\label{sec:smooth_gauge}

The topological invariant defined in Eq.~\eqref{eq:invariant}
requires a continuous gauge. In the following, we are going to present a method
to construct such a gauge from numerically obtained eigenstates of a 1D system.
The discussion closely follows the appendix of Ref.~\onlinecite{SoV12}.

Let us consider an isolated set of $N$ bands, i.e., the bands can cross each
other but shall have no crossings with other bands outside the considered set.
Furthermore,
consider a discrete uniform $k$ mesh of $M+1$ points $\lbrace k_j\rbrace$,
$j\in[1,M+1]$, where $k_{j+1}=k_j + \Delta$ and $k_{M+1}=k_1 + G$ with a
reciprocal lattice vector $G$. We label the corresponding eigenstates along the
mesh $|\tilde u_{nk_j}\rangle$, where $n$ is the band index. If these states are
obtained from a numerical diagonalization routine, they will in general have
random phases. It is easy to see that in the limit $M\rightarrow \infty$ such a
choice of phases is highly non-differentiable. 

In order to construct a smooth numerical gauge, we have to define what we mean
by ``smooth'' for a discrete mesh. This can be done by requiring that the states
remain as parallel as possible as we move along a path from $k_1$ to $k_{N+1}$.
In other words, the change in the states should be orthogonal to the states
themselves. The corresponding gauge is called \emph{parallel transport gauge}.
For a single isolated band this can be realized by choosing the phases of the
Bloch states such that the overlap $\langle u_{nk_j}|u_{nk_{j+1}}\rangle$ is
real and positive. For $N$ bands, one has to require that the overlap matrix
$L_{mn}=\langle u_{mk_j}|u_{nk_{j+1}}\rangle$ is hermitian with only positive
eigenvalues.

We are now going to explain how a parallel-transport gauge can be constructed in
practice. We start from the initial point $j=1$ where we set
$|u_{nk_1}'\rangle=|\tilde u_{nk_1}\rangle$. Then, at each subsequent $k_{j+1}$
we have to rotate the states $|\tilde{u}_{nk_{j+1}}\rangle$ by a unitary matrix
$U$ in such a way that the overlap matrix $\tilde{L}$ becomes hermitian and
positive. This is accomplished by employing a singular value decomposition. More
specifically, $\tilde{L}$ can be written as $\tilde{L}=V\Sigma W^\dagger$, where
$V$ and $W$ are unitary and $\Sigma$ is positive real diagonal. If we now set
$U=WV^\dagger$ and transform the states as
\begin{equation}
|u_{nk_{j+1}}'\rangle= \sum_m^N U_{mn}(k_{j+1})|\tilde{u}_{mk_{j+1}}\rangle,
\end{equation} 
the new overlap matrix becomes $L'=V\Sigma V^\dagger$, which is hermitian and
positive. Repeating this for the entire $k$ mesh, we finally get a set of states
$|u_{nk_j}'\rangle$ that are smooth in the sense specified above. However,
states at $k_1$ and $k_{N+1}$ will in general differ by a unitary transformation
$\Lambda$ and are, thus, not mapped onto themselves via parallel transport.
The matrix $\Lambda$ corresponds to a non-Abelian analog of the Berry
phase and has eigenvalues of the form $\lambda_l=e^{i\phi_l}$. 

The periodicity can be restored in the following way: we first determine the
unitary matrix $S$ that diagonalizes $\Lambda$. We then rotate all states at all
$k_j$ by $S$. This results in a gauge in which the new states correspond to a
diagonal $\Lambda$ with eigenvalues $e^{i\phi_l}$. Finally, we spread the
residual phase differences over the $k$ mesh,
\begin{equation}
|u_{lk_j}\rangle = e^{i(j-1)\phi_l/M} |u_{lk_j}'\rangle.
\end{equation}
Eventually, we have constructed a smooth, periodic gauge for the set of $N$
bands. Note, however, that the new states $|u_{lk_j}\rangle$ are in general not
eigenstates of the Hamiltonian. Nevertheless, at each $k_j$ they span the
eigenspace corresponding to the $N$ bands. They can therefore be used for the
calculation of $U(N)$-invariant quantities like the partial polarization.

\end{document}